\documentclass[12pt]{article}
\usepackage{amsfonts,amssymb,amsmath,latexsym}
\newcommand{\be}{\begin{equation}}
\newcommand{\ee}{\end{equation}}
\newcommand{\bea}{\begin{eqnarray}}
\newcommand{\eea}{\end{eqnarray}}

\newcommand{\ba}{\begin{array}}
\newcommand{\ea}{\end{array}}

\newcommand{\bt}{\begin{tabular}}
\newcommand{\et}{\end{tabular}}

\newcommand{\fr}{\frac}
\newcommand{\ci}{\cite}
\newcommand{\cl}{\centerline}
\newcommand{\bs}{\bigskip}
\newcommand{\vs}{\vspace}

\newcommand{\en}{\eqno}

\newcommand{\bbib}{\begin{thebibliography}}
\newcommand{\ebib}{\end{thebibliography}}

\newcommand{\mbb}{\mathbb}

\newcommand{\und}{\underline}

\begin{document}

\titlepage
\hspace{5cm} {\it LANDAU INSTITUTE preprint 06/01/00}
\vspace{1.5cm}

\centerline{\bf 3D VAN DER WAALS}

\centerline{\bf $\sigma$-MODEL AND  ITS TOPOLOGICAL EXCITATIONS}

%\cl{\la{\bf WITH LOGARITHMIC ENERGY}}

\vs{0.5cm}

\cl{\bf S.A.Bulgadaev}

%\cl{\bf S.A.Bulgadaev\fm{}}\fnt{This paper is supported by
%RFBR grants 96-02-17331-a and 96-15-96861}

\bs

\cl{\it L.D.Landau Institute, Kosyghin Str.2, Moscow, Russia,117334}
%\cl{E-mail:bulgad@itp.ac.ru}
\vs{0.5cm}

%\cl{Abstract}

\vs{0.4cm}

It is shown that 3D vector van der Waals (conformal) nonlinear
$\sigma$-model (NSM) on a sphere $S^2$ has two types of
topological excitations reminiscent vortices and instantons of 2D NSM.
The first, the hedgehogs, are described by homotopic group
$\pi_2(S^2) = \mbb {Z}$ and have logarithmic energies. They are an analog
of 2D vortices. The energy and interaction of these
excitations are found. The second, corresponding to 2D instantons, are
described by homotopic group $\pi_3(S^2) = \mbb {Z}$ or the Hopf invariant
$H \in \mbb {Z}$.
A possibility of the topological
phase transition in this model and its
applications are briefly discussed.

\bs

\cl{PACS: 11.15.Kc, 11.27.+d, 61.30.Jf, 75.10.Hk}

\newpage

% Plan

%I.Intro
%1)Importance of t.e., their interrelations with symmetry and correlations
%a) shrinked and open boundary,
%b) conformal symmetry and its breaking, role of log-exitations in PT in
%low-dim. systems
%c) a possibility to go to D>2
%II. 3D-dimensional generalization, 3D case on $S^2$
%1.Conformal $\sigma$-models, Fourier space, nonlocal action, regularization,
%Green function, its log behaviour.
%Simple case of $S^{d-1}$, equations, their geometry,
%III. vortex-instanton with log energy
%1) behaviour of vector field with unit norm,
%2) gauss form, top.solutions, step-approximation, fourier space representation
%3)energy, log-divergence, role of reg.function
%IV.Discussion.
%1. Usual approach, mean-field, phase transition, simple estimation
%2.Instantons-vortices
%4.Factorization, curvature, interaction of different modes
%renormalization of spin waves
%5.Effective theory, duality
%6.Renormalization, low T approach, break of conformal sym.
%7.Local theory and possible properties
%IV. Applications
%1. Coulomb interaction
%2. Vectorial pi and top.charges, F_G
%3. Possible aplications

%\cl{\bf I. Introduction.}

As is known, the topologically stable excitations
can exist in systems with degenerate minima, which form some manifold
${\cal M}$ with  nontrivial topology.
The topological  excitations (TE) take place in many systems
and play important role in determination of the physical properties of
these systems. But this influence strongly depends on interrelations
between topology and symmetry of the systems. In the most interesting cases
of scale invariant systems the TE
determine a behaviour of correlations in systems due to dynamical scale
symmetry breaking and appearence of the effective mass scale
$$
m \sim a\exp (-{\cal S}_{TE}),
\en(1)
$$
($a$ is an effective "frequency" or an UV cut-off parameter) connected with
a dimensionless "energy" (or an action) ${\cal S}_{TE}$
of the TE \ci{1}. The corresponding scale invariant low-dimensional
($D \le 2$) models
are used for description of many important physical phenomena such as
coherence-decoherence, localization and topological phase transitions \ci{2,4}.
For example, it is the
vortices in 2D $XY$-model or non-linear $\sigma$-model (NSM) on a
circle $S^1$ \ci{3,4},
and the kinks in 1D long-range Ising model \ci{5} and in quantum
dissipative models \ci{2,6}, what determines the correlation of these models.
Moreover,
a phase transition (PT) in the system of TE changes the correlation
of the whole system [2-7].
In general, not all TE can  have such strong influence on correlations.
In 2D NSM on a sphere $S^2$ there
are also the TE, the instantons \ci{8}, which give a finite contribution to
the mass generation. But analogous
mass generation takes place in all NSM
on any compact  spaces with non zero
curvature (or Ricci tensor), though not all of them have the TE \ci{9}.
The different influence of the TE in these models is connected with different
properties of their TE. The vortices and kinks in above models
interact with each other
through logarithmic potential, while the instantons do not interact
between themselves, and only more weak, a dipole-dipole like, interaction
exist between instantons and anti-instantons \ci{9}.
It is just a logarithmic interaction what induces a topological PT (TPT) in
low-dimensional with $D \le 2$ systems \ci{4,5,7}.

For this reason it is interesting to find  the higher dimensional models
($D \ge 2$),
having the TE with logarithmic energy, and to define a role of these TE.

The necessary conditions for existence of both these TE in D dimensions are
a nontriviality of the homotopic groups $\pi_{D-1}({\cal M})$   and
$\pi_D({\cal M})$ respectively.
A simple analysis shows that a logarithmic
divergence of the energy of the TE corresponding to nontrivial abelian group
$\pi_{D-1}({\cal M})$
is determined by the two next properties:

1) they must correspond to the open boundary $S^{d-1}$ of space
$\mbb {R}^d$;

2) a scale (or even conformal) invariance of the model action.

A sphere $S^2$  satisfies all these conditions in 3D space $\mbb {R}^3$.
For this
reason we confine ourselves in this paper by the case
${\cal M} = S^2$.

Let us consider 3D simple cubic lattice with the order parameter (OP)
${\bf n}$ in each lattice site, taking its values in $S^2$ or in $RP^2.$
The corresponding OP can be:

1) a unit vector ${\bf n}, \quad n^2 =1, \quad {\bf n} \in S^2,$
it can represent a magnet;

2) a unit rod or a director  ${\bf n} \in S^2/Z_2 = RP^2,$
it can represent a liquid crystal or molecular crystal.

Since the relevant homotopy groups of $S^2$ and $RP^2$ are identical \ci{10}
$$
\pi_2(S^2) = \pi_2(RP^2) = \mbb {Z},
\en(2)
$$
$$
\pi_3(S^2) = \pi_3(RP^2) = \mbb {Z},
\en(3)
$$
it will be more convenient to consider the OP ${\bf n} \in S^2.$
All results will takes place with non-essential modifications for
${\bf n} \in RP^2$  too.
Due to its vectorness,  the OP can interact by different type
of interactions:

1) exchange type  $\sim ({\bf n}_r \cdot {\bf n}_{r'}) V(r-r')$,
where $V(r)$ can be short or long-range,

2) dipole-like  $\sim  (n_r^i D_{ik}(r-r') n_{r'}^k )/r^3, \quad
D_{ik}(r) = \delta_{ik} - 3\fr{x^i x^k}{r^2}$

All they can be represented in the next form
$$
{\cal E} \sim (n^i_r D_{ik} n^k_{r'}) V(r-r'),\quad
V(r) \sim 1/r^{\sigma},
\en(4)
$$
where  $\sigma$ defines an asymptotic behaviour of the potential $V(r).$

As a first approximate attempt to the problem one can compose from all
these types of interaction the simplified one, which must conserve
the two main properties:

1) a scale invariance of the corresponding Hamiltonian ${\cal H},$

2) a vectorness of the OP.

Since in 3D the van der Waals asymptotics
$$
V(r) \sim 1/r^6,
\en(5)
$$
ensures a scale invariance of ${\cal H}$ on large scales,
one gets in result

\noindent \und {the lattice vector van der Waals model} with ${\cal H}$
$$
{\cal H} = - \fr{J}{2}
\sum_{r \ne r'} ({\bf n}_r \cdot {\bf n}_{r'})V_{vdW}(r-r').
\en(6)
$$
The analogous approximation, for example, was used by Nelson \ci{11}
in the theory of 2D melting.

In long-wave continuous approximation our model passes into the

\noindent \und {vector van der Waals NS-model} with a partition function
$$
{\cal Z}_{vdW} =\int D{\bf n} e^{- {\cal S}_{vdW}[{\bf n}]},
\en(7)
$$
$$
{\cal S}_{vdW}[{\bf n}] =  - \fr{1}{2\alpha}\int d^3x d^3x'
({\bf n}(x){\bf n}(x'))V_{vdW}(x-x'),
\en(8)
$$
$$
V_{vdW}(x)= \left.\int \fr{d^3k}{(2\pi)^3} e^{i({\bf k}{\bf x})}|k|^3 f(ka)
\right|_{|x| \gg a}
\sim 1/|x|^6
\en(9)
$$
where
$${\bf n}^2 =1, \quad \alpha \sim  1/J\beta,
\en(10)
$$
$f(ka)$ is a regularizing function with next asymptotics
$$
f(ka)_{ka \ll 1} \simeq 1+O(ka), \quad
f(ka)_{ka \gg 1} \to 0.
\en(11)
$$
From now on we omit an index vdW for brevity.
A scale invariance of the model at large distancies follows
immediately from large-distance asymptotics of $V(x)$ and
dimensionlessness of the OP ${\bf n}.$ Moreover, ${\cal S}$ is
conformal invariant at large distancies, i.e. it is invariant
under conformal transformations:
$$
x_i \to  x'_i = x_i/r^2, \quad r \to r' = 1/r, \quad x_i/r = x'_i/r',
\en(12)
$$
$$
d^3 x \to d^3 x/|{\bf x}|^6,\quad
\fr{1}{|{\bf x}_1-{\bf x}_2|^{6}} \to
\fr{|{\bf x}_1|^6 \;|{\bf x}_2|^6}{|{\bf x}_1-{\bf x}_2|^{6}},
\en(13)
$$
and, consequently,
$$
{\cal S} \sim \int d^3 x_1 d^3 x_2 \;
\fr{({\bf n}_1{\bf n}_2)}{|{\bf x}_1-{\bf x}_2|^{6}} \to {\cal S}
$$
For this reason  this model can be  named also the
\und {3D conformal NS-model} \ci{12}.

The corresponding Euler - Lagrange equation has a form
$$
\int V(x-x') {\bf n}(x')d^3 x' -
{\bf n}(x) \int ({\bf n}(x){\bf n}(x')) V(x-x') d^3 x' = 0.
\en(14)
$$
The Green function $G(x)$ of the conformal kernel
$V(x)$ can be defined by next equation
$$
\int V(x-x'') G(x''-x') d^3 x'' = \delta (x-x')
\en(15)
$$
It has the following form
$$
G(x) = \left.\int \fr{d^3 k}{(2\pi)^3}\fr{e^{i({\bf k}{\bf x})}}
{k^3 f(ka)} \right|_{r \gg a} \simeq
- \fr{1}{(2\pi)^{3/2} (2)^{1/2}
\Gamma(3/2)} \ln (r/R)
\en(16)
$$
and a logarithmic asymptotic behaviour.

One can
consider the action (8) from very beginning as a part of a more general,
non scale invariant, model with an action ${\cal S}_g$,
including a local gradient term
${\cal S}_l$,
$$
{\cal S}_g = {\cal S} + {\cal S}_l, \quad
{\cal S}_l = \fr{1}{2A} \int d^3x (\partial {\bf n})^2.
\en(17)
$$
Its form in Fourier space will be
$$
{\cal S}_g = \fr{1}{2} \int \fr {d^3 k}{(2\pi)^3}
|n(k)|^2 k^2 \left( \fr{1}{A} +  \fr{k}{\alpha} + O(k^2)
\right),
\en(18)
$$
A similar action was obtained earlier in the theory of liquid crystals
by taking into account interaction between fluctuations \ci{13}.
If in the system there is a point,
where a local "rigidity" $\fr{1}{A}= 0$,
then in this point one obtains  a scale invariant action (8) with a
first term $\sim k^3$. From a point of view of the Ginzburg-Landau theory
this case is similar to the
\und {tricritical point}, where a term $\sim \psi^4$ is absent in the
expansion of effective potential of the theory on nonlinearities \ci{14}
$$
V(\psi) = a \psi^2 + b \psi^4 + c \psi^6.
\en(19)
$$
This expansion in 3D is also non scale invariant, but at the tricritical
point $a=b=0,$ and a theory becomes scale invariant since a term
$c \psi^6$ is scale invariant \ci{14}.

An usual analysis (see for example \ci{14}) shows that a model (8)
does not have a usual phase transition with nonzero OP due  to logarithmic
divergence of the OP fluctuations
$$
(\delta {\bf n}(x))^2 \sim \int d^3 k G(k) \sim \int d^3 k /k^3 \sim \ln (R/a),
\en(20)
$$
where $\delta {\bf n}(x)$ is a deviation from some fixed value ${\bf n}_0,$
$R$ and $a$ are, respectively, the size of system and a short-range
cut-off parameter. In this sense a 3D vector van der Waals NSM (8)
is analogous to 2D XY-model \ci{3,4}.
The thermodynamic properties of the model (8) will be
considered in other paper \ci{15}. Here we will study only its possible TE.

It is convenient to write equation (14) in a "linear" form,
introducing new function $g(x):$
$$
\int V(x-x') {\bf n}(x') d^3 x' = g(x) {\bf n}(x),
\en(21)
$$
$$
g(x) = \int ({\bf n}(x){\bf n}(x')) V(x-x') d^3 x'.
\en(22)
$$
As all equations for NSM on spheres, it means that the action of the
operator $V$ on vector ${\bf n}(x)$ must be proportional to this vector, i.e.
the vector field ${\bf n}(x)$ must be in some sence an "eigenvector"
of the operator $V$ with
the "eigenvalue" $g(x)$, functionally depending on ${\bf n}(x).$
For each solution of equation (21) a value of the ${\cal S}[{\bf n}]$ can
be expressed through the "eigenvalue"
$$
{\cal S} = \fr{1}{2\alpha} \int g(x) d^3 x
\en(23)
$$
Since $\pi_2(S^2)= \mbb {Z}$, there are the TE with topological charge
$Q \in \mbb {Z}$.
It is worthwhile to mention here that in 3D case there is strong
difference between
the tangent and normal to some closed surface vector fields in contrast
with 2D case. In the latter case both the vortices and the hedgehog
excitations can exist. They have the same
logarithmic energy. In the 3D case there is the well known constraint
on such tangent vector fields, so called "no hair-dressing" theorem, stating
that  these tangent vector fields must have the singularities \ci{10}.
For this reason one needs to consider only the hedgehog type solutions.
The simplest TE with charge $Q=1,$ corresponding to the
identical map of spheres $S^2,$ has the next asymptotic form
$$
{\bf n}(x)_{r \gg a} \simeq  \fr{x^i}{r}
\en(24)
$$
Substituting (24) into equation (21), passing to the momentum space and using
the Fourier-image of the corresponding function
$$
n^i(k) = -  \fr{8\pi i}{k^3} \quad \fr{k^i}{k},
\en(25)
$$
we see, after returning to the $x$-space,
that the field (24) is an "eigenvector" of $V$ and, consequently,
a solution of equation (21).
The corresponding "eigenvalue" is
$$
g(x) = \fr{8}{(2\pi)^{1/2}  r^3}
\int dk k^{1/2} J_{1/2}(k) f(ka/r) = \fr{8}{\pi r^3}.
\en(26)
$$
The action ${\cal S}$ of this solution is
$$
{\cal S} \simeq \fr{(8\pi)^2}{2\alpha} \fr{4\pi}{(2\pi)^3}
\int \fr{dk}{k} f(ka)
\en(27)
$$
The integral in (27) is logarithmically divergent as it should be.
After introducing a radius $R$ of the space and using asymptotic behaviour
of function $f(ka)$ it can be written in
the next form
$$
\int_{1/R}^{\infty} \fr{dk}{k} f(ka) =
\ln (R/a) - \int \ln k \; df(k).
\en(28)
$$
If $f(k)$ has a sharp, a step-like form $f(k) = 1- \theta(k-1)$  (what means
a lattice regularization), then
an integral in (28) is zero and we get a pure logarithmic action
$$
{\cal S} = \fr{C}{\alpha} \ln (R/a), \quad
C = \fr{(8\pi)^2}{2} \fr{4\pi}{(2\pi)^3} = 16.
\en(29)
$$
One can show that the interaction of two different
TE with charges
$Q_1$ and $Q_2$ on large distancies
has a form of the Green function $G(r)$
$$
H_{12} (r) = Q_1 Q_2 G(r) \simeq  -  \fr{Q_1 Q_2}{(2\pi)^{3/2} (2)^{1/2}
\Gamma(3/2)} \ln (r/R)
\en(30)
$$
Note that in the usual local 3D NS-model (17) such TE have the energy linear
in $R$
$$
E \simeq \fr{4\pi}{A} (R-a).
\en(31)
$$
It is interesting that if we consider the full action (18),
then the corresponding
equation has again the "hedgehog" solution (24) with total energy
$$
E = \fr{4\pi}{A}(R-a) + \fr{16}{\alpha} \ln (R/a).
\en(32)
$$
It means that a logarithmic part of the "hedgehog" energy can be
observed also in general model (18) at scales
$$
A/\alpha > l > a.
$$
It is clear that analogous "anti-hedgehog" solutions exist in these models
with the same energy.

A nontriviality of another homotopical
group $\pi_3(S^2) = \mbb {Z}$ means that the "neutral"
configurations, having a full topological charge $Q = 0$ and
an asymptotic behaviour corresponding to the shrinked boundary,
$$
{\bf n}(r) = {\bf n}_0 ,\quad  r \to \infty, \quad  \mbb {R}^3 \to S^3,
\en(33)
$$
can also have different topological structures.
They are characterized by topological invariant, coinciding with
the Hopf invariant $H \in \mbb {Z}$
of the corresponding mapping $S^3 \to S^2.$
This invariant is connected  with linking number and can be expressed
through the integrals over $\mbb {R}^3$ \ci{10}
$$
\{\gamma_1, \gamma_2 \} =
\fr{1}{4\pi} \oint_{\gamma_1} \oint_{\gamma_2}
\fr{<{\bf r}_{12} \cdot [d{\bf r}_1 d {\bf r}_2]>}
{|r_1-r_2|^3}.
$$
For simple case of one winding of one circle around another $H = 1.$
If a mapping projects each circle $q_i \;(i=1,2)$ times then $H= q_1 q_2.$

In $S^3$ it can be also represented in a local form as
$$
H = \int_{S^3} \theta \wedge d \theta,
$$
where 1-form $\theta$ is defined as
$$
d \theta = f^{-1}( d\Omega).
$$
Here $d \Omega$ is a 2-form or an element of the area of $S^2,$
$f^{-1}$ is a mapping  inverse to the  projection mapping.
Some particular case of its explicit form
can be found in \ci{16}.
Just this invariant is an analog of the topological charge of 2D instantons.
Note, that this additional topological invariant classifies "neutral"
(relative to the group $\pi_2(S^2)$)
configurations in all 3D NS-models defined on sphere $S^2,$ in particular,
in the usual local (17) and in general (18) models. In this relation
one can conjecture that just this invariant
defines anomaly strong slowness of the "hedgehog-anti-hedgehog"
recombination rate in recent experiments in liquid crystals \ci{17}.
This takes place due to existence of some energetical barriers between
"neutral" configurations corresponding to different Hopf invariant $H.$

Thus, the "hedgehog" excitations in 3D van der Waals NSM  have
properties reminiscent of the mixed properties of the two-dimensional
vortices and instantons:

1) their topology is described by $\pi_2(S^2)$,
but they interact as vortices through logarithmic potential;

2) their "neutral" configurations are classified by integer topological
Hopf invariant $H$.

In principle, the TE with logarithmic energy can induce TPT in system
with such TE.
The simple arguments by Kosterlitz and Thouless \ci{3,4} show this for
any dimensional case, giving a critical temperature $T_{KT}$
$$
T_{KT} \simeq \fr{\beta}{\alpha} C_D/(D k_B),
\en(34)
$$
where $\beta = 1/T, \, C_D$ is a corresponding coefficient in
the logarithmic energy,
$k_B$ is the Boltzmann constant.
It means that at  $T > T_{KT}$ it becomes energetically favorable
to birth such TE and they can generate spontaneously, while at
$T < T_{KT}$ one needs positive free energy to birth these TE. More
detailed study of this transition in low-dimensional systems has shown
that these arguments correspond to the first order approximation in
the renorm-group (RG) approach to the dilute gas of TE approximation
of the initial NSM.  In higher orders of RG  the contributions,
taking into account more detailed information about geometry of space
${\cal M},$ type of the corresponding topological charges and
their interaction, appear in RG equations.

In 3D van der Waals NS-model there are some additional complications,
they will be considered in paper \ci{15}.

The van der Waals NSM, by its construction, is a simplified model
of real systems, consisting from the rod-like molecules, interacting
through the van der Waals potential. But, since
topological characteristics depend on rough
qualitative properties, not on some inessential details,
one can hope that
the proposed van der Waals model can describe the qualitative
properties of some real systems.

An application of ideas, developed under
investigation of this model, to the more
realistic, taking into account an anisotropy of the liquid crystals,
model from \ci{13} may be  especially interesting.
The possible generalizations of this consideration on other ${\cal M}$
and dimensions $D$ are  discussed in \ci{12}.

\bigskip

The author is very obliged to H.Brand, E.Kats, M.Kleman, M.Monastyrs-kii
and many others for usefull and fruitfull discussions.

This work was supported by RFFI grant 96-15-96861.
A part of this work was  done during a stay at the
Max Plank Institute
for Physics of Complex System, Dresden, Germany. The author is thankfull
to the administration of institute for hospitality and support.
My thanks also to the organizers of the workshop
"Topological Defects in Non-Equilibrium Systems and Condenced Matter" for
the opportunity to give a talk.

\end{document}